\documentclass{aa_new}

\usepackage{graphicx}

\newcommand{\Ho}{H_{\rm o}}

\newcommand{\lamo}{\lambda_{\rm o}}

\newcommand{\OmT}{\Omega_{\rm tot}}

\begin{document}

%
   \title{Bias in  Matter Power Spectra ?}
%

        \titlerunning{Bias in Matter Power Spectra ?}

   \author{M.~Douspis$^{1,3}$, A.~Blanchard$^{1,2}$ \& J.~Silk$^3$}
   \authorrunning{M.~Douspis, A.~Blanchard \& J.~Silk}

   \offprints{}

   \institute{$^1$ Observatoire Midi-Pyr\'en\'ees,
              14, ave. E. Belin,
              31400 Toulouse, FRANCE \\
              Unit\'e associ\'ee au CNRS, UMR 5572\\
              $^2$ Universit\'e Louis Pasteur,
                   4, rue Blaise Pascal, 
                   67000 Strasbourg,
                  FRANCE\\
	      $^3$ Astrophysics, Nuclear and Astrophysics Laboratory,
                   Keble Road,
		   Oxford, OX1 3RH,
	           UNITED KINGDOM\\
             }

   \date{Mai 2001}

   \abstract{
We review the constraints given by the  linear matter power
spectra data on cosmological and bias parameters,
comparing the data from the 
PSCz survey (Hamilton et al., 2000) and from the matter power spectrum 
infered by 
the study of Lyman alpha spectra at z=2.72 (Croft et al., 2000).
We consider flat--$\Lambda$ cosmologies, allowing $\Lambda$, $H_0$ and $n$ 
to vary, and we also let the two  ratio factors $r_{pscz}$ and  
$r_{lyman}$ ($r^2_i = \frac{P_{i}(k)}{P_{CMB}(k)}$) vary independently.  
Using a simple $\chi^2$ minimisation 
technique, we find confidence 
intervals on our parameters for each dataset and for a combined analysis.
Letting the 5 parameters vary freely  gives almost no constraints on cosmology,
but requirement of a universal ratio
for both datasets implies unacceptably  low values of $H_0$
 and $\Lambda$. Adding some reasonable
priors on the cosmological parameters
demonstrates  that the power derived by the PSCz 
survey  is higher by a factor $\sim 1.75$ compared to the power from the Lyman 
 $\alpha$ forest survey.
      \keywords{matter power spectrum -- lyman alpha -- Cosmology: 
observations -- Cosmology: theory}
}
   
	\maketitle


\section{Introduction}

Bias, defined to be the ratio of rms luminosity density fluctuations to 
dark matter fluctuations at a fiducial scale of 12$h_{2/3}^{-1}$ Mpc, is 
the most uncertain of the minimal set of cosmological parameters.
These may be taken to be   $\Lambda$, $H_0$, $\Omega_0$ and
 $\Omega_b$ for the background, with the inhomogeneities described by
 $n$ and the bias parameter $b$. 

There are three techniques available for computing the 
shape and amplitude of the power spectrum of density
fluctuations. These involve using the CMB anisotropy data (BOOMERanG, MAXIMA, DASI), galaxy redshift surveys (PSCz, Hamilton \& Tegmark, 2000) and the Lyman alpha forest (Croft et al., 2000).  Use of large-scale peculiar velocity
surveys and of galaxy cluster abundances provide a value of the normalisation, via the parameter  $\sigma_8\Omega_0^{0.6}$, where  $\sigma_8\equiv b$ 
evaluated at the fiducial scale of 12$h_{2/3}^{-1}$ Mpc.
We confirm that CMB and galaxy surveys provide a consistent data set 
for the power spectrum. However incorporation of the matter power spectra 
inferred from Lyman alpha forest requires a relative bias of galaxies 
relative to the Lyman alpha forest
once reasonable priors are adopted for the cosmological parameters.

\begin{figure}
\begin{center}
\resizebox{\hsize}{!}{\includegraphics[angle=0,totalheight=8.4cm,
        width=8.cm]{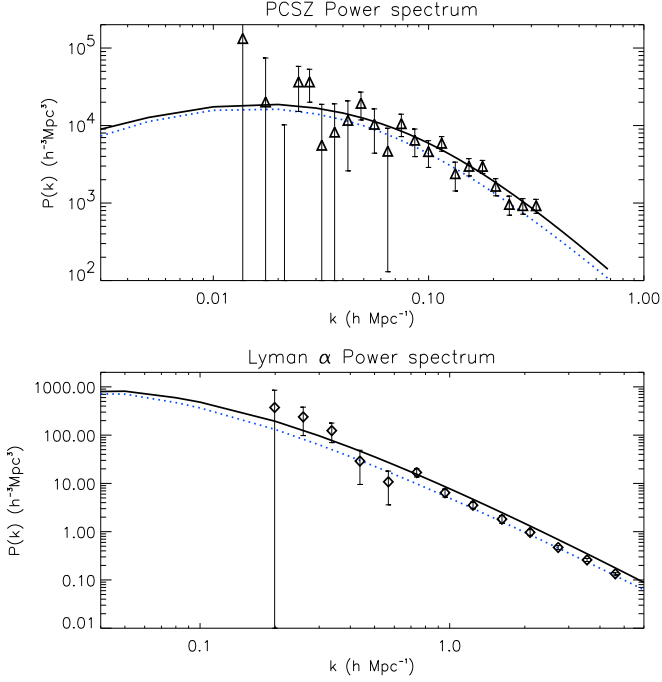}}
\end{center}
\caption{\label{fig_powplot1}Actual data set of matter power spectra. Up: PSCz spectrum. Down: Lyman alpha power spectrum. The solid line is the best model of the combined analysis (see text) and the (blue) dotted line correspond to the best model of the CMB, PSCz, Lyman $\alpha$ combined analysis.}

\end{figure}

\section{Data sets}

\subsection{PSCz}

The linear power spectrum inferred by the PSCz survey is taken from Hamilton 
\& Tegmark 2000 (HT00 hereafter). We choose the 22 points given in table B2
 of HT00,
 which correspond to estimates of the decorrelated linear power spectrum. 
The dataset is shown in the upper part of figure \ref{fig_powplot1}.

\subsection{Lyman $\alpha$}

We use the linear matter power spectrum inferred from the flux power
 spectrum of 
lyman $\alpha$ forest spectra by Croft et al., 2000 (C00 hereafter). This 
corresponds to
13 estimates of the spectrum at $<z>=2.72$, referenced in table 4 of C00.
When these data are  used, the normalisation uncertainty
(+27\%, -23\%) is included,
 then marginalised over (see next paragraph). 
The data are drawn in the lower part of figure \ref{fig_powplot1}.

\section{Cosmologies}

\subsection{Free parameters}

In this paper, we consider flat cosmologies, with  no reionisation, no 
massive neutrinos and no tensor fluctuations. We also fix the baryon fraction to be consistent with Big--Bang nucleosynthesis and CMB determinations:
$\Omega_b h^2=0.020$. The remaining parameters are 
listed in table 1, as well as  the corresponding ranges and steps. All the 
theoretical power spectra are COBE--normalised, and the ratio parameters
are defined as:
$$P(k)_{pscz/lyman} = r_{pscz/lyman}^2 P_{cmb}(k).$$
Here $\lamo$ is the vacuum density and $n$ is the fluctuation spectral 
index.

\begin{table}[h]
\begin{center}
\begin{tabular}{|c|c|c|c|c|c|c|c|}
\hline
 & $\Ho\ kms^{-1}/Mpc$&$\lamo$&$n$& $r_{pscz}$& $r_{lyman}$ \\
\hline
\hline
Min. & 20  & 0.0   & 0.70  & 0.50& 0.10 \\
\hline
Max. & 100 & 1.0   & 1.30  & 1.475& 2.05 \\
\hline
step & 10  & 0.1   & 0.03 & 0.025& 0.05  \\
\hline
\end{tabular}
\end{center}
\caption{\label{tab1}Parameter space explored:} 
\vspace{-0.3cm}
$\OmT = 1$\\
$\Omega_b h^2 = 0.020$ 
\end{table}

\subsection{Theoretical matter power spectrum}

We have considered theoretical matter power spectra with an arbitrary value 
of the  shape
parameter ($\Gamma$), using the approach of 
\cite{bardeen} and \cite{sugiyama}. We
computed the COBE normalisation of the spectra using the CAMB (\cite{camb}) 
code. 
We used the ``grow package'' from \cite{hamilton} to 
derive the growth
factor. Our two theoretical power spectra at z=0 and z=2.7 are
presented in the following equations:
\begin{eqnarray}\label{eqPp}
P_{z=0}(k)  &=& N_1 \times TF(\Gamma,k)^2 \times g(\Lambda)^2 \times P_o(k) *r_{pscz}^2\\
P_{z=2.72}(k)  &=& N_2 \times f_z \times TF(\Gamma,k)^2 \times g(\Lambda)^2 \times P_o(k) *r_{lyman}^2 
\end{eqnarray}
where we  write:
\begin{equation}
f_z = (1+z)^{-2}*[(\Lambda+\Omega_m*(1.+z)^3+ \Omega_k*(1+z)^2)^{1/2}]^3
\end{equation}
and the initial matter power spectrum is given by:
\begin{equation}
P_o(k) = (k*h)^n
\end{equation}
We take  the transfer function to be:
\begin{eqnarray}
& &TF(\Gamma, k) = \frac{ln(1+2.34*q)}{(2.34*q)}\times\nonumber\\
& &\frac{1}{(1+\alpha_2*q+(\alpha_3*q)^2+(\alpha_4*q)^3+(\alpha_5*q)^4)^{1/4}}
\end{eqnarray}
where $q=k/\Gamma$. We use the coefficients given by  Sugiyama (1995): $\alpha_1=2.34,\ \alpha_2=3.89,\ \alpha_3=16.1,\ \alpha_4=5.46,\ \alpha_5=6.719$.
Here we treat $\Gamma$ as a free parameter.
We then marginalize over it. The shape of the power spectrum will
be investigated in  a future paper.
$N_1$ and $N_2$ are normalisation factors, such that  our matter power 
spectra are COBE--normalised.

\section{Statistics}

\subsection{Likelihood}

We approximate the likelihood functions of our parameters by independent multivariate 
Gaussians,  using the following expression:
$$\mathcal{L}_{pscz/lyman} = exp -\frac{\chi^2_{pscz/lyman}}{2}$$
For the PSCz data, the $\chi^2$ has been computed as follows:
\begin{equation}\label{eq1}
\chi^2_{pscz}(\Theta_p) = \sum \frac{(P_i(k_i) - P^{model}(\Theta_p|k_i))^2}{\sigma_i^2.}
\end{equation}
where $P^{model}$ is derived from equation \ref{eqPp} of the following section
 and $\Theta_p$ is the set of parameters explored listed in table \ref{tab1}.

Concerning the Lyman alpha data, we take into consideration the uncertainty
 on the normalisation. We add a contribution to  $\chi^2$, assuming
 a double--tail Gaussian for the normalisation distribution:
\begin{eqnarray}\label{eq2}
\hat{\chi}^2_{lyman}(\Theta_l, N) = \sum \frac{(P_i(k_i) - P(\Theta_l, N|k_i)^2}{\sigma_i^2.}\nonumber\\
 + \frac{(N-1.)^2.}{(\sigma^2)}\nonumber\\
\chi^2_{lyman}(\Theta_l) \equiv -2 \times \log \left[\int exp(-\frac{\hat{\chi}^2_{lyman}}{2}) dN\right]
\end{eqnarray}

\subsection{Results}

The results are shown in 2D contours plots, with red-dashed lines 
corresponding to confidence intervals 68, 95, 99\% in one dimension,
 and filled contours corresponding to 
68, 95, 99\% in two dimensions.
The remaining parameters have been marginalised (not integrated), 
finding the minimum for each pair of parameters plotted in the 
graphs.

\section{Results and constraints}

\subsection{PSCz and Lyman $\alpha$ matter spectra}

Given our 5 free parameters, the constraints given by each set independently 
are almost inexistent. Many combinations of parameters lead to degeneracies 
which allow values even out of our grid of parameters.

\subsection{Are the two data sets consistent ?}

We have seen that our two data sets allow most of the values for the
parameters we are considering. Nevertheless, this does not mean that
they are compatible. Marginalising over parameters, for a better
view of the constraints, does not allow us to include all the
correlations and degeneracies between parameters. This is even more
critical when the constraints are weak. To see whether the data are
consistent in some particular cosmologies, we adopt two scenario. The
first is to consider that the ratios are the same for both
datasets. Marginalising over these will reveal   the prefered range
for the cosmological parameters. The second approach is to let 
 the ratios vary freely, while
putting some priors on the
cosmological parameters. This will give us  estimates of the  ratio for
each data set.

	\subsubsection{Uniform ratio}

\begin{figure}
\begin{center}
\resizebox{\hsize}{!}{\includegraphics[angle=0,
        totalheight=7cm,
        width=7.cm]{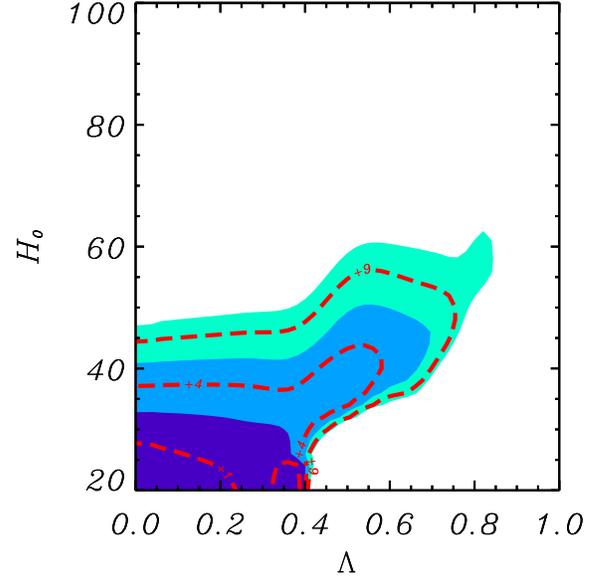}}
\end{center}
\caption{\label{figpclyb}Combined constraints from the Lyman $\alpha$
power spectrum and the PSCz spectrum with the same ratio. Filled contours mark
the 68, 95 and 99\% confidence levels in 2--dimensional space. The dashed red 
line define the 68, 95 and 99\% confidence levels when projected on 1 parameter.}
\end{figure}

We have combined the  data sets by adding the two $\chi^2$ grids
defined in equations \ref{eq1} and \ref{eq2}, $r$ being the unique bias
parameter. The prefered  models  obtained with this assumption correspond
to very low values of $H_o$ and $\Lambda$. The ``best one'' is plotted in 
figure \ref{fig_powplot1}.
The value of
the goodness of fit for this model, given by the absolute value of the
$\chi^2$ on the two dataset, is good ($\chi^2_{min} = 22$ for 30
degrees of freedom\footnote{The degrees of freedom are given by the
number of points in the $\chi^2$ minus the number of free parameters
investigated; $dof=N_{pts}-N_{param}$. This is actually correct only
if the $\chi^2$ is linearly dependent on the parameters, but it is
generally used, and  still gives  a good estimate of the goodness of
fit.}).   The
corresponding confidence intervals are shown in figure \ref{figpclyb},
where the remaining parameters have been marginalised over. This
combined analysis give strong constraints on the cosmological
parameters, which are   largely in contradiction with most  estimates
of the
cosmological parameters  from independent techniques (such as SNIa, age of the Unviverse, CMB).  Although the concordance model is marginally acceptable,  a low $H_0$ tilted CDM model is preferred.

	\subsubsection{Ratio parameters}	

 The preceding discussion  does not reinforce our
 faith in  the concordance model with respect to  other parameter
estimates. The
analysis of the data could  be done via an alternative approach.  Assuming
that we ``know'' the cosmological parameters, we may examine   the relative
 bias of the two
data sets. Statistically speaking, this means that we can put priors on
the cosmological parameters, marginalise over them, and see what the
constraints are on the 2 remaining parameters that are associated with the
 inhomogeneities power. We adopt the reasonable 
following priors in our analysis: $\Lambda=0.7 \pm 0.1$, $H_0 = 65 \pm 10\
kms^{-1}Mpc^{-1}$, $n=1\pm0.1$ which are in agreement with most of the actual
parameter estimates in the literature. 
We assume a Gaussian shape for the
priors, giving the following likelihood:
\begin{eqnarray}
\hat{\chi}^2 &=& \chi^2_{pscz} + \chi^2_{lyman} + \chi^2(\Lambda) +
\chi^2(H_0) + \chi^2(n)\nonumber\\  
\chi^2(\Lambda) &=& \frac{(\Lambda-0.7)^2}{0.1^2}\nonumber\\
\chi^2(H_0) &=& \frac{(H_0-65)^2}{10^2}\nonumber\\
\chi^2(n) &=& \frac{(n-1.0)^2}{0.1^2}\nonumber
\end{eqnarray}

then
\begin{equation}
\chi^2_{combined}(r_{pscz}, r_{lyman}) = Min_{[\Lambda, H_0, n]}(\hat{\chi}^2)
\end{equation}

\begin{figure}
\begin{center}
\resizebox{\hsize}{!}{\includegraphics[angle=0]{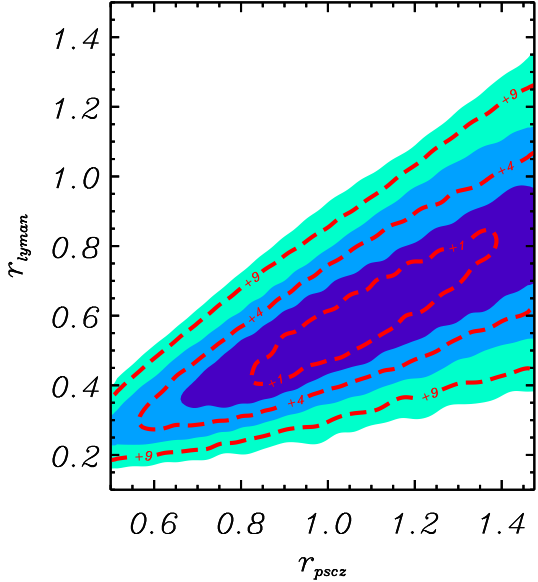}}
\end{center}
\caption{\label{figpclyBB}Combined Constraints from the Lyman $\alpha$
matter power spectrum and the PSCz spectrum with priors on cosmological
parameters (see text)}

\end{figure}

Figure \ref{figpclyBB} shows the confidence intervals 
for $r_{pscz}$ and $r_{lyman}$. The best model is obtained for
$r_{pscz}=1.075$ and $r_{lyman}=0.6$, with a good goodness
 of fit:
$\chi^2=24.8$ for 29 degrees of freedom. We can clearly see, in the figure 
\ref{figpclyBB} the correlation
between the two biases that leads to a relation (over the explored range of 
parameters and between the 95\% confidence level interval):
\begin{equation}
r_{pscz} = 1.57*r_{lyman}+0.16\;\;\; or \;\;\; \frac{r_{pscz}}{r_{lyman}}\sim1.8
\end{equation}
The corresponding 68\% intervals on each ratio are: $0.45 < r_{lyman} < 0.8$ and $0.8 < r_{lyman} < 1.35$; the constraints being low at 95\% CL.

\subsubsection{Combining CMB and matter spectra}

\begin{figure}
\begin{center}
\resizebox{\hsize}{!}{\includegraphics[angle=0,totalheight=7cm,
        width=7.cm]{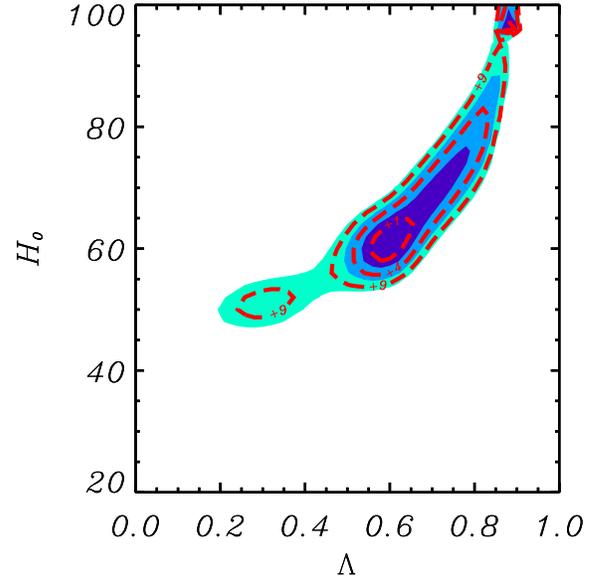}}
\end{center}
\caption{\label{figcmb}CMB constraints using COBE, BOOMERanG, DASI and MAXIMA datasets for flat $\Omega_bh^2=0.02$ models.}
\end{figure}  

\begin{figure}
\begin{center}
\resizebox{\hsize}{!}{\includegraphics[angle=0]{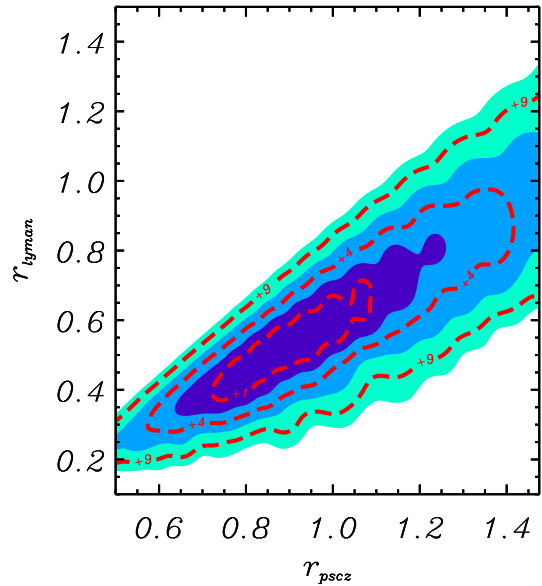}}
\end{center}
\caption{\label{figCPbvb}Combined Constraints from CMB, PSCz spectrum
and the Lyman $\alpha$ power spectrum in the ($r_{pscz}, r_{lyman}$) plane}
\end{figure}  

Instead of setting independent priors on each cosmological parameter,
we can use the latest CMB analysis as a prior. In LeDour et al., 2000
and  Douspis et al., 2001, we used the current CMB data set to derive
constraints on cosmological parameters. The latter paper considers the most
recent data including BOOMERanG (Netterfield et al., 2001), DASI (Halverson et  al., 2001) and  MAXIMA (Lee et al., 2001).   We use the same
grid as in table 1 for the CMB analysis. As seen in the literature
(Douspis 2000, Jaffe et al., 2001, Melchiorri and Griffiths, 2000 and others), 
the CMB by itself does not constrain $H_0$ or $\Lambda$; both of
these parameters  are  degenerated with the total density (or the curvature). 
As
we have fixed $\Omega_b h^2 = 0.02$ and $\Omega_{total}=1$ in this
analysis, the CMB  now gives interesting confidence intervals on the
remaining 
cosmological parameters (see figure \ref{figcmb}). These constraints are 
similar to those used
in the previous section, but contain more information, because of the
correlations between parameters. This way of setting priors should
 be preferable to the previous one. 

We constructed a new likelihood grid from the CMB and the matter power
spectra:
\begin{eqnarray}
\chi^2(\theta)&=&\mathcal{A}_{cmb} + \chi^2_{pscz} +\chi^2_{lyman}\\
\rm with\; \; \; \theta &=& (\Lambda, H_0, n, \Gamma, r_{pscz}, r_{lyman}) 
\end{eqnarray}
where $\mathcal{A}$ is the approximation developed in Bartlett et al., 2000.
We then maximise this new grid and marginalise to find confidence intervals.
The best model given by $\Lambda=0.6,\ H_0=60,\ n=0.97,\ \ r_{pscz}=0.875,\ r_{lyman}=0.5$ has a good goodness of fit and
 is plotted in figure \ref{fig_powplot1}.

As the constraints on cosmological parameters are stronger in the CMB analysis
than in the matter power spectrum analysis, we do not expect any improvement
on cosmological parameter confidence intervals. On the other hand, 
marginalising over $\Lambda,\ H_0,\ n$ gives us interesting constraints on 
$r_{pscz}\ {\rm and} \ r_{lyman}$. When the three
 datasets are combined, we  find constraints on the relative bias and again 
 a degenerated relation between the two
 ratio parameters, leading to a lower ratio factor for the Lyman linear power
 spectrum (cf figure  \ref{figCPbvb}). The 68 \% confidence level are the
 following: $0.45 < r_{lyman} < 0.6$ and $0.7 < r_{lyman} < 1.1$. 
We may also express the relative bias between the two sets of data as:
\begin{equation}
r_{pscz} = 1.33*r_{lyman}+0.20\;\;\; or\;\;\; \frac{r_{pscz}}{r_{lyman}}\sim 1.75
\end{equation}
inside the 95\% confidence levels.

\section{Conclusions}

Determination of the  cosmological model parameters is entering into
a new era of precision. The weakest link in our understanding of the universe is in the realm of the density fluctuations, and their evolution.
It is well known that bias depends on the objects being sampled, 
and varies with
galaxy luminosity and morphological type. However these variations
are attributed to the complex physics of galaxy formation.
We find  here that by adopting reasonable cosmological model priors, 
the matter power spectrum inferred by Lyman alpha forest power spectrum
 can be shown to be 
 biased low  relative to the
the bias of $L_\ast$ galaxies. 

It is usually assumed that the Lyman alpha forest, representing gas that has not yet virialized and has density contrast between unity and 200, is a good tracer of the dark matter (e.g. Croft et al. 2001). In fact this is a dangerous assumption,
since the Lyman alpha forest gas is a very small fraction of the total gas density, 99.99 percent or more of which is ionized. Moreover, the
proximity effect observed for quasars and more recently for Lyman break galaxies (Steidel 2001), demonstrates that the forest is not a good tracer of 
the total matter density on scales of up to
tens of megaparsecs.

We note that gas-rich dwarf galaxies, which avoid the vicinities of 
luminous galaxies, also display a significantly lower bias, or even
 an antibias, compared to that of  luminous galaxies. Dwarfs are likely 
to be the final fate of much of the forest.
Indeed LCDM simulations of clustering
find that luminous galaxies themselves are somewhat antibiased.
Hence our  result fits in well with hierarchical
structure formation,
 since the low Lyman alpha  forest  ratio $\frac{r_{pscz}}{r_{lyman}}\sim 1.75$
is similar
 to that of dwarf galaxies.




\begin{thebibliography}{}

\bibitem[Bardeen 1986]{bardeen} Bardeen et al., 1986, ApJ, 304, 15
\bibitem[Bartlett, Douspis, Blanchard, \& Le 
Dour(2000)]{2000A&AS..146..507B} Bartlett, J.\ G., Douspis, M., Blanchard, 
A., \& Le Dour, M.\ 2000, A\&AS, 146, 507 
\bibitem[Croft et al. 2000]{lyman} Croft, R.A.C. et al., astro-ph/0012324
\bibitem[Douspis 2000]{douspis} Douspis, M., 2000, PhD thesis
\bibitem[Douspis et al  2001]{douspis et al} Douspis M.,  Blanchard A., Sadat R.,  Bartlett, J.G.,\& Le Dour M., A\&A in press, astro-ph/0105129
\bibitem[]{} Halverson et al., astro--ph/0104489 (DASI)
\bibitem[Hamilton 2001 ]{hamilton} Hamilton, A.\ J.\ S.\ 2001, MNRAS, 322, 419
\bibitem[Hamilton \& Tegmark 2000]{pscz} Hamilton, A.J.S. \& Tegmark, M., astro-ph/0008392
\bibitem[Jaffe et al 2001]{jaffe}  Jaffe et al. 2001, Phys. Rev. Lett., 86, 3475-3479
\bibitem[Le Dour, Douspis, Bartlett, \& 
Blanchard(2000)]{2000A&A...364..369L} Le Dour, M., Douspis, M., Bartlett, 
J.\ G., \& Blanchard, A.\ 2000, A\&A, 364, 369 
\bibitem[]{} Lee et al., astro--ph/0104459 (MAXIMA)
\bibitem[Lewis, Challinor, \& Lasenby 2000]{camb} Lewis, 
A., Challinor, A., \& Lasenby, A.\ 2000, ApJ, 538, 473 
\bibitem[]{}Melchiorri A. \& Griffiths L.M., New Astronomy Reviews, 45, Issue 4-5, 2001
\bibitem[]{} Netterfield et al., astro--ph/0104460 (BOOMERanG)
\bibitem[Seljak \& Zaldarriaga 1996]{cmbfast} Seljak U. \& Zaldarriaga M. 1996, ApJ 469, 437
\bibitem[]{}Steidel C., 2001, private communication
\bibitem[Sugiyama 1995]{sugiyama} Sugiyama, N.\ 1995, ApJS, 
100, 281 



\end{thebibliography}
\end{document}